\documentclass[aps,showpacs,pra,superscriptaddress]{revtex4}
\usepackage{amsmath}
\usepackage{graphicx}
\usepackage{subfigure}
\begin{document}

\title{Visibility of  ultracold Bose system in triangular optical lattices }
\author{Zhi Lin}
\affiliation{Department of Physics, Shanghai University, Shanghai
200444, P.R. China}
\author{Jun Zhang}
\affiliation{Department of Physics, Shanghai University, Shanghai
200444, P.R. China}
\author{Ying Jiang}
\thanks{Corresponding author}
\email{yjiang@shu.edu.cn} \affiliation{Department of Physics,
Shanghai University, Shanghai 200444, P.R. China} \affiliation{Key
Lab for Astrophysics, Shanghai 200234, P.R. China}

\begin{abstract}

In this paper, by treating the hopping parameter in Bose-Hubbard
model as a perturbation, with the help of the re-summed Green's
function method and cumulants expansion, the momentum distribution
function of the ultra-cold Bose system in triangular optical
lattice is calculated analytically. By utilizing it, the
time-of-flight absorption picture is plotted and the corresponding
visibility is determined. The comparison between our analytical
results and the experimental data from
Ref.[\onlinecite{sengstock}] exhibits a qualitative agreement.

\end{abstract}


\maketitle

\section{Introduction}

During the past decade, the properties of ultra-cold atomic gases
in optical lattices \cite{Markus-Greiner} have received a great
deal of attention on account of their novelty and various
potential applications \cite{Immanuel-Bloch}. In the case of
bosons, the delicate balance between the atom-atom on-site
repulsion and the hopping parameter leads to a Mott
insulator-Superfluid (MI-SF) quantum phase transition which has
been confirmed experimentally \cite{Markus-Greiner}. Due to the
precisely controllable parameters, the
optical-lattice-ultra-cold-atom systems play roles of test ground
for quantum many-body systems in condensed matter physics, this is
the so-called quantum simulation \cite{lewenstein-1}.

Beside the simple cubic optical lattices, in which almost all
experiments with ultra-cold atoms to date have been performed due
to the ease of the ease of the implementation, the triangular and
hexagonal optical lattices have also been recognized recently
\cite{sengstock}. In fact, due to the complexity of the lattice
structure in these systems, novel and rich new phases will be
exhibited, for instance, the effects of geometrical frustration
have been achieved in triangular optical lattice cold atomic
system experimentally \cite{sengstock-science,Eckardt}. Hence, it
is worth to investigate these systems, especially their quantum
phase transitions, in a systematic way.

In our previous work \cite{jiang-pra}, we have calculated the
phase boundaries of MI-SF quantum phase transitions of ultra-cold
bosons in triangular, hexagonal, as well as Kagom\'e optical
lattices analytically via the method of the field-theoretical
effective potential \cite{axel}, the relative deviation of our
analytical results from the numerical results
\cite{numerical-results} is less than 10\%. However, these results
cannot be compared with the experimental observation, since the
phase boundaries of the ultra-cold system are not able to be
detected directly in experiments. Instead, time-of-flight
measurement \cite{Immanuel-Bloch}, which reveals the momentum
distribution of the system, is the standard experimental technique
for investigating cold atom systems in optical lattices
\cite{Markus-Greiner1,Michael}.

In this paper, by treating the hopping parameter as a
perturbation, with the help of Green's function calculation
\cite{Metzner,ohliger}, we are going to calculate the
time-of-flight absorbtion pictures and the associated visibility
as well analytically for ultra-cold scalar bosons in triangular
optical lattice. Our results exhibit a qualitative agreement with
the experimental data.

\section{The model}

As is well known, a system of scalar bosons trapped in a
homogeneous optical lattice is portrayed by the many-body
Hamiltonian which takes the form as
\begin{equation}
H=\int d{\bf r}
\psi^{\dagger}(\textbf{r})\left(-\frac{\hbar^2}{2m}{\nabla^2}
+V(\textbf{r})\right)\psi(\textbf{r})+\frac{g}{2}\int d\textbf{r}
\psi^{\dagger}(\textbf{r})\psi^{\dagger}(\textbf{r})\psi(\textbf{r})\psi(\textbf{r}),
\label{original-hamiltonian}
\end{equation}
with $\psi(\textbf{r})$ ($\psi^{\dagger}(\textbf{r})$) being
annihilation (creation) bosonic field operators and
$V(\textbf{r})$ being the optical lattice trapping potential. The
parameter $g$ is the interaction strength between two atomic
particles.

In fact, due to the modulation of the optical lattice potential
wells, in the extremely low temperature limit, a single band
approximation is adequate, the bosonic field operators
$\psi^{\dagger}(\textbf{r})$\ and $\psi(\textbf{r})$ can be
expanded in the basis of orthonormal Wannier functions
$w(\textbf{r}-\textbf{r}_{j})$ of the lowest band
\cite{jaksch-1,blakie} as
\begin{eqnarray}
\psi(\textbf{r})&=&\sum_{j}\emph{w}\,(\textbf{r}-\textbf{r}_{j})\hat{a}_{j},
\nonumber \\
\psi^{\dagger}(\textbf{r})&=&\sum_{i}\emph{w}^{\ast}\,(\textbf{r}-\textbf{r}_{i})\hat{a}_{i}^{\dagger},
\label{field-operator-expansion}
\end{eqnarray}
where $\hat{a}_{i}^{\dagger}$ ($\hat{a}_{i}$) is the creation
(annihilation) operator of scalar boson on site of $i$.

By utilizing the above expressions, the Hamiltonian in
Eq.(\ref{original-hamiltonian}) is reduced in the tight-binding
limit to the Bose-Hubbard Hamiltonian
\begin{equation}
\hat{H}_{\rm BH}=\hat{H}_{1}+\hat{H}_{0} \label{equation1}
\end{equation}
with the diagonal part $\hat{H}_{0}$ and the hopping part
$\hat{H}_{1}$ being
\begin{equation}
\hat{H}_{0}=\sum_{i}\frac{U}{2}\hat{n}_{i}(\hat{n}_{i}-1)-\mu\hat{n}_{i},\,\,\,\hat{H}_{1}=-J\sum_{\langle
i,j\rangle}\hat{a}_{i}^{\dagger}\hat{a}_{j}, \label{equation2}
\end{equation}
respectively, where $\hat{n}_{i}=
\hat{a}_{i}^{\dagger}\hat{a}_{i}$\ is the particle number operator
on site $i$. $J$ is the hopping amplitude for the bosons between
the nearest neighbor sites $i$ and $j$ while $U$ denotes the
on-site repulsion and $\mu$ is the chemical potential. As we can
see from Eq.(\ref{original-hamiltonian}) to Eq.(\ref{equation2}),
$J$ and $U$ can be calculated explicitly by \cite{jaksch-1}
\begin{equation}
J=-\int d{\bf r}w^*({\bf r}-{\bf
r}_i)\left(-\frac{\hbar^2\nabla^2}{2m}+V({\bf r})\right)w({\bf
r}-{\bf r}_j) \label{J-formula}
\end{equation}
and
\begin{equation}
U=g\int d{\bf r}|w({\bf r})|^4, \label{U-formula}
\end{equation}
respectively, with ${\bf r}_i$ and ${\bf r}_j$ being nearest
neighboring sites.

In experiment, the triangular optical lattice is created by three
laser beams intersecting in the $x$-$y$ plane with angle of
$2\pi/3$ between each other \cite{sengstock}, the lattice constant
of the obtained triangular lattice is $a=2\lambda/3$, where
$\lambda$ is the wavelength of the lattice laser beams. The
corresponding lattice trapping potential reads
\begin{equation}
V({\bf
r})=V_0\left\{\frac34-\frac12\left[\cos\left(\frac{4\pi}{\sqrt{3}a}y\right)
+ \cos\left(\frac{2\pi}{\sqrt{3}a}y-\frac{2\pi}{a}x\right) +
\cos\left(\frac{2\pi}{\sqrt{3}a}y+\frac{2\pi}{a}x\right)\right]\right\}.
\label{lattice-potential}
\end{equation}
Associated to this optical lattice trapping potential, in harmonic
approximation, the corresponding Wannier function takes form of
\begin{equation}
w({\bf r}-{\bf
r}_i)=\left(\frac{32\widetilde{V}_0}{9}\right)^{\frac14}
\left(\frac{\pi}{a^2}\right)^{\frac12}\exp\left[-\left(\frac{8\widetilde{V}_0}{9}\right)^{\frac12}
\frac{\pi^2}{a^2}\left((x-x_i)^2+(y-y_i)^2\right)\right]
\label{wannierfunction}
\end{equation}
with $\widetilde{V}_{0}$ being dimensionless optical lattice depth
in unit of recoil energy $E_R=(\hbar^2 k^2)/2m$, where
$k=2\pi/\lambda$.

\section{The momentum distribution function}

Actually, the quantity that is measured in experiments is the
distribution function in the momentum space \cite{prokofev-1}
\begin{equation}
n(\mathbf k)=\int d\mathbf r\,d\mathbf r^{\prime} e^{i\mathbf
k\cdot(\bf r-\bf r^{\prime})} \langle
\psi^{\dagger}(\bf{r})\psi(\bf{r^{\prime}})\rangle.
\label{equation3}
\end{equation}
By making use of Eq.(\ref{field-operator-expansion}), with the
help of the Fourier transformation of the operators
\begin{equation}
\hat{a}^{\dagger}_i=\frac {1}{\sqrt{N_S}}\sum_{\mathbf
k_{1}}\hat{a}^{\dagger}_{\mathbf k_{1}}e^{-i\mathbf k_{1}
\cdot\mathbf r_{i}},\,\,\,\hat{a}_j=\frac
{1}{\sqrt{N_S}}\sum_{\mathbf k_{2}}\hat{a}_{\mathbf
k_{2}}e^{i\mathbf k_{2}\cdot\mathbf r_{j}}, \label{equation6}
\end{equation}
where $N_S$ is the total of the lattice number, the momentum
distribution function is reduced to
\begin{equation}
n(\mathbf k)=N_S\mid\emph{w}(\mathbf k)\mid^2\langle
\hat{a}^{\dagger}_{\mathbf k} \hat{a}_{\mathbf k} \rangle.
\label{equation7}
\end{equation}

As is known, the one particle Green's function in the momentum
space reads
\begin{equation}
G(\tau \mid 0, \mathbf k)=\,\langle
\hat{T}_{\tau}[\hat{a}^{\dagger}(\tau)_{\mathbf k}
\hat{a}(0)_{\mathbf k}] \rangle,
\end{equation}
where $\tau$ is imaginary time and $\hat{T}_{\tau}$ is the
imaginary  time ordering operator. It is the Fourier
transformation of the one particle Green's function $
G(\tau^{\prime},j^{\prime}\mid \tau,j)=\,\langle \hat{T}_{\tau}
[\hat{a}^{\dagger}_{j^{\prime}}(\tau^{\prime}) \hat{a}_j(\tau)]
\rangle $. Apparently, $\langle \hat{a}^{\dagger}_{\mathbf k}
\hat{a}_{\mathbf k} \rangle = \lim_{\tau\downarrow 0}G_{1}(\tau
\mid0,\mathbf k)$, we then translate the problem of calculating
the density distribution to a problem of calculating corresponding
Green's function.

However, due to the non-commutativity of the two parts of the full
Hamiltonian (\ref{equation1}), the exact information of the
corresponding eigenstates can hardly be obtained. Nevertheless,
the Green's functions can be calculated perturbatively via the
Dirac representation. To tackle the issue of time-of-flight
absorbtion picture which is closely related to the MI-SF phase
transition, we treat the hopping term in the Hamiltonian
(\ref{equation1}) as the perturbation part, for the opposite limit
will not lead to a MI-SF phase transition at all \cite{stoof}.

In Dirac picture, the time evolution of operators is only
determined by the unperturbed part of the Hamiltonian (set
$\hbar=1$) \cite{M. Peskin} $
\hat{O}(\tau)=e^{\tau\hat{H}_{0}}\hat{O}e^{- \tau\hat{H}_{0}}$,
and the corresponding evolution operator takes the form of
\begin{equation}
\hat{u}(\beta,0)=\hat{T}_{\tau}\left[e^{\int^{\beta}_{0}d\tau\sum_{\langle
i,j \rangle} J
\hat{a}^{\dagger}_{i}(\tau)\hat{a}_{j}(\tau)}\right].
\label{equation10}
\end{equation}
Without any difficulty, it can be proved that the one particle
Green's function reads
\begin{eqnarray}
G(\tau^{\prime},j^{\prime}\mid
\tau,j)&=&\,\frac{\text{Tr}\{{{e^{-\beta
\hat{H}_{0}}}\hat{T}_{\tau}[\hat{a}^{\dagger}_{j^{\prime}}(\tau^{\prime})\hat{a}_{j}(\tau)\hat{u}(\beta,0)]}\}}
{\{\text{Tr}{e^{-\beta \hat{H}_{0}}}\hat{u}(\beta,0)\}}.
\label{equation40}
\end{eqnarray}
To calculate $G(\tau^{\prime},j^{\prime}\mid \tau,j)$, the
evolution operator $\hat{u}(\beta,0)$ has to be expanded
perturbatively, i.e., the expansion express of
$G(\tau^{\prime},j^{\prime}\mid \tau,j)$ consists of terms, for
instance, as
\begin{eqnarray}
\frac{1}{n!}\sum_{i_{1},j_{1},\cdots,i_{n},j_{n}}J_{i_{1},j_{1}}\cdots
J_{i_{n},j_{n}}\int^{\beta}_{0}d\tau_{1} \cdots
\int^{\beta}_{0}d\tau_{n}
\langle\hat{T}_{\tau}[\hat{a}^{\dagger}_{j^{\prime}}(\tau^{\prime})\hat{a}_{j}(\tau)\hat{a}^{\dagger}_{i_{1}}(\tau_{1})
\hat{a}_{j_{1}}(\tau_{1})\cdots\hat{a}^{\dagger}_{i_{n}}(\tau_{n})\hat{a}_{j_{n}}(\tau_{n})]\rangle_{0},
\end{eqnarray}
with $\langle \hat{O}\rangle_{0}$ being average quantity with
respect to the unperturbed part $H_0$ of the Hamiltonian. Here,
the $J_{ij}$ reads
\begin{equation}J_{ij}=\begin{cases}
 J, \;\;\;& \text{if {\it i, j} are nearest neighbors of each other,}\\
 0, \;\;\;&\text{otherwise.}
\end{cases}
\end{equation}
Thus, the following quantities need to be calculated first
\begin{equation}\label{equation11}
G^{(0)}_{n}(\tau^{\prime}_{1},i^{\prime}_{1};\cdots\tau^{\prime}_{n},i^{\prime}_{n}\mid
\tau_{1},i_{1};\cdots\tau_{n},i_{n}) =\langle
\hat{T}_{\tau}[\hat{a}^{\dagger}_{i^{\prime}_{1}}(\tau^{\prime}_{1})\hat{a}_{i_{1}}(\tau_{1})\cdots
\hat{a}^{\dagger}_{i^{\prime}_{n}}(\tau^{\prime}_{n})\hat{a}_{i_{n}}(\tau_{n})]\rangle_{0},
\end{equation}
these are the $n$-particle Green's function with respect to $H_0$.
Unfortunately, the Wick's theorem cannot be applied to calculate
the above quantity, since the eigenstates of the non-perturbed
part of the Hamiltonian (\ref{equation1}) is local, and each
annihilation operator in the expression should be paired with a
creation operator on the same site, otherwise the result would be
zero. Therefore, we should turn to the theory of linked-cluster
expansion \cite{Metzner,ohliger}, i.e. to expand the $n$-particle
Green's function in terms of the cumulants
$C^{(0)}_m(\tau^{\prime}_{1},\cdots,\tau^{\prime}_{m}\mid\tau_{1},\cdots\tau_{m})
=\langle
\hat{T}_{\tau}[\hat{a}^{\dagger}(\tau^{\prime}_{1})\hat{a}(\tau_{1})\cdots
\hat{a}^{\dagger}(\tau^{\prime}_{m})\hat{a}(\tau_{m})]\rangle_{0}$
in which the particle operators are all on the same site.

By defining the generation function as
\begin{equation}\label{equation12}
C^{(0)}_{0}[j,j^{\ast}]=
\ln\langle\hat{T}_{\tau}e^{(\int^{\beta}_{0}d\tau[j^{\ast}(\tau)\hat{a}(\tau)+j(\tau)\hat{a}^{\dagger}(\tau)])}\rangle_0,
\end{equation}
all the cumulants can then be calculated by
\begin{equation}\label{equation13}
C^{(0)}_{m}(\tau^{\prime}_{1},\cdots,\tau^{\prime}_{m}\mid\tau_{1},\cdots\tau_{m})=\left.\frac{\delta^{2m}}{\delta
j(\tau^{\prime}_{1})\cdots\delta j(\tau^{\prime}_{m})\delta
j^{\ast}(\tau_{1})\cdots\delta
j^{\ast}(\tau_{m})}C^{(0)}_{0}[j,j^{\ast}]\right|_{j=j^{\ast}=0}
\end{equation}
These cumulants can be used to decompose the above mentioned
$n$-particle Green's function. As an example, the decomposition of
one and two particle Green's functions is shown in the following
\begin{equation}\label{equation14}
G^{(0)}_{1}(\tau^{\prime},i^{\prime}\mid\tau,i)=\delta_{i^{\prime},i}\,C^{(0)}_{1}(\tau^{\prime}\mid
\tau)
\end{equation}
\begin{eqnarray}\label{equation15}
G^{(0)}_{2}(\tau^{\prime}_{1},i^{\prime}_{1};\tau^{\prime}_{2},i^{\prime}_{2}\mid
\tau_{1},i_{1};\tau_{1},i_{1})&=&
\delta_{i^{\prime}_{1},i_{1}}\delta_{i^{\prime}_{1},i^{\prime}_{2}}\delta_{i_{1},i_{2}}\,C^{(0)}_{2}(\tau^{\prime}_{1},\tau^{\prime}_{2}\mid \tau_{1},\tau_{2})\nonumber \\
& &+\delta_{i^{\prime}_{1},i_{1}}\delta_{i^{\prime}_{2},i_{2}}C^{(0)}_{1}(\tau^{\prime}_{1}\mid \tau_{1})C^{(0)}_{1}(\tau^{\prime}_{2}\mid \tau_{2})\nonumber \\
& &+\delta_{i^{\prime}_{1},i_{2}}\delta_{i^{\prime}_{2},i_{1}}C^{(0)}_{1}(\tau^{\prime}_{1}\mid \tau_{2})C^{(0)}_{1}(\tau^{\prime}_{2}\mid \tau_{1})
\end{eqnarray}

In order to reduce the complication of the calculation, these
cumulants can be represented diagrammatically as
\begin{equation}\label{equation16}
C^{(0)}_{1}(\tau^{\prime}\mid
\tau)=\raisebox{-0.4cm}{\includegraphics[width=2cm]{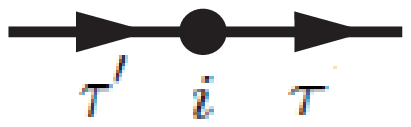}},
\end{equation}
\begin{equation}\label{equation17}
C^{(0)}_{2}(\tau^{\prime}_{1},\tau^{\prime}_{2}\mid
\tau_{1},\tau_{2})=\raisebox{-0.7cm}{\includegraphics[width=2cm]{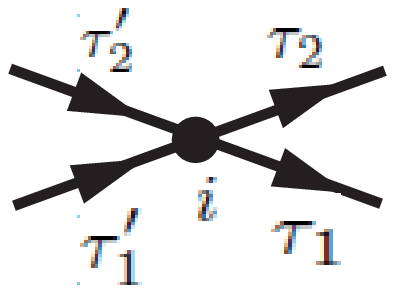}},
\end{equation}
meanwhile, the hopping parameter $J_{ij}$ is represented
diagrammatically as
\begin{equation}\label{equation16-1}
J_{ij}=\raisebox{-0.3cm}{\includegraphics[width=2cm]{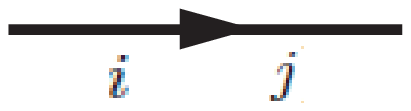}}.
\end{equation}
With the help of the cumulants and the diagram rules mentioned
above, each term in the expansion expression of
$G(\tau^{\prime},j^{\prime}\mid \tau,j)$ can then be represented
diagrammatically easily. When considering the cancellation effect
of the denominator in Eq.(\ref{equation40}), the one particle
Green's function $G(\tau^{\prime},j^{\prime}\mid \tau,j)$ only
consists of connected diagrams.

In perturbative calculation, the choice of the diagrams of the
Green's function is very subtle. In the problem of investigating
the phenomena closely related to the SF-MI phase transitions,
choosing terms based on the order of $J/U$ is not a good
approximation, since in the regime of superfluid, this quantity is
no longer a small quantity \cite{Axel2}, more importantly, the
Green's function should be diverging around the phase transition
point, however it is impossible to obtain such diverging behavior
from a finite-order perturbation calculation, since it yields only
a polynomial of $J/U$. Instead, we consider a resummed Green's
function\cite{ohliger} which contains only the diagrams of single
chains, in terms of Matsubara frequency, it is expressed as
\begin{equation}\label{equation18}
G_{1}(\omega_{m},i,j)=\raisebox{-0.42cm}{\includegraphics[width=2cm]{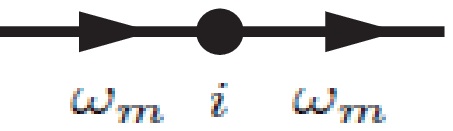}}+
\raisebox{-0.33cm}{\includegraphics[width=2.7cm]{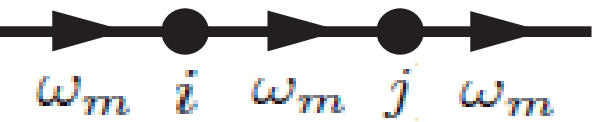}}+\cdots.
\end{equation}
The reason for such a choice is following. Let us compare these
two terms
\begin{equation}
\includegraphics[width=1.5cm]{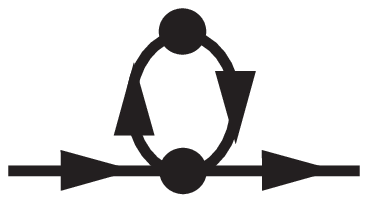}\;\;,\;\;\;\;\includegraphics[width=3.3cm]{5-1.eps},
\end{equation}
although both of them are second order terms of hopping parameter,
however, according to the topology of the underlying lattice
structure, the former is smaller than the latter by factor of
$\frac1{d}$, $d$ is the dimension of the system. In general, if a
diagram contains $n$ loops, it will be smaller than the
corresponding single-chain diagram at least by factor of
$\frac{1}{d^n}$. Similar argument was stated in the case of
fermions \cite{Metzner}. In the limit of $d\rightarrow \infty$,
comparing to single-chain diagrams, all diagrams with loops may be
neglected. In this sense, our choice of the resummed Green's
function can be looked upon as a sort of mean-field treatment.
However, from the above discussion, we see that, in principle, the
Green's function method can easily be extended to a regime beyond
mean-field.

When translating this diagram expression to the expression in
terms of cumulants and hopping parameters, through the Fourier
transformation, we get
\begin{equation}\label{equation19}
G_{1}(\omega_{m},\mathbf k)= \sum^{\infty}_{l=0}\left[C^{(0)}_{1}(\omega_{m})\right]^{l+1} (J(\mathbf k))^{l}
\end{equation}
where $J(\mathbf k)$ is Fourier transformation of the $J_{ij}$
\begin{equation}\label{equation20}
J(\mathbf k )= \sum _{ij} J_{ij} e^{i \mathbf k\cdot\mathbf
r_{i}}e^{- i\mathbf k\cdot\mathbf r_{j}}
\end{equation}
and
\begin{equation}\label{equation21}
C^{(0)}_{1}(\omega_{m})=\int^{\beta}_{0}C^{(0)}_{1}(\tau)e^{i\omega_{m}\tau} d\tau
\end{equation}
In the case of  the triangle lattice, $J(\mathbf k )$ reads
\begin{equation}\label{equation22}
J(\mathbf k )= 2J\left[\cos(k_{x}a)+\cos(k_{x}\frac
{a}{2}+k_{y}\frac {\sqrt{3}a}{2})+\cos(k_{y}\frac
{\sqrt{3}a}{2}-k_{x}\frac {a}{2})\right].
\end{equation}
Since
\begin{equation}
C^{(0)}_{1} (\tau)=\,\langle
\hat{T}_{\tau}[\hat{a}^{\dagger}(\tau) \hat{a}(0)] \rangle_{0},
\label{equation 24}
\end{equation}
together with $Z^{(0)}={\rm Tr}[e^{-\beta H_0}]$ and
$H_{0}|n\rangle= E_{n}|n\rangle$ ($|n\rangle$ stands for a state
with $n$ particles on the site), we then have
\begin{equation}\label{equation25}
C^{(0)}_{1}( \tau)=\frac{1}{Z^{(0)}}\sum^{\infty}_{n=0}
\left[\theta(\tau)n e^{(E_{n}-E_{n-1})\tau}+\theta(-\tau)
(n+1)e^{-(E_{n}-E_{n+1})\tau}\right]e^{-\beta E_{n}}.
\end{equation}
In terms of Matsubara frequency, it reads
\begin{equation}\label{equation26}
C^{(0)}_{1}(\omega_{m})=\frac{1}{Z^{(0)}}\sum^{\infty}_{n=0}\left[\frac{n+1}{E_{n+1}-E_{n}+i\omega_{m}}-\frac{n}{E_{n}-E_{n-1}+i\omega_{m}}
\right]e^{-\beta E_{n}}.
\end{equation}

Since the temperature in experiment is extremely low, thermal
fluctuations are negligible \cite{Axel2}. In the limit of
$T\rightarrow 0$, the system of unperturbed part falls into the
ground state (suppose the occupation number in the ground state
being $n$), and the momentum space Green's Function is then
reduced to
\begin{eqnarray}
G_{1}(\tau^{\prime}\mid0,\mathbf
k)&=&\frac{1}{2\pi}\sum^{\infty}_{l=0}\int^{\infty}_{-
\infty}\left[\frac{n+1}{E_{n+1}-E_{n}+i\omega_{m}}-\frac{n}{E_{n}-E_{n-1}+i\omega_{m}}\right]^{l+1}(J(\mathbf
k))^{l}e^{- i\omega_{m}\tau^{\prime}}d\omega_{m}\nonumber \\
&=&\frac{1}{2\pi}\sum^{\infty}_{l=0}\int^{\infty}_{-
\infty}\left(\frac{J(\mathbf
k)}{U}\right)^{l}\left[\frac{-1-\widetilde{\mu}+i\widetilde{\omega}_{m}}
{(n-\widetilde{\mu}+i\widetilde{\omega}_{m})(n-1-\widetilde{\mu}+i\widetilde{\omega}_{m})}\right]^{l+1}e^{-
i\widetilde{\omega}_{m}\widetilde{\tau^{\prime}}}d\widetilde{\omega}_{m}
\label{eqnarray 29}
\end{eqnarray}
where $\widetilde{\mu}=\frac{\mu}{U}$,
$\widetilde{\omega}_{m}=\frac{\omega_{m}}{U}$,
$\widetilde{\tau^{\prime}}=\tau^{\prime}U$. After a laborious yet
straightforward calculation, we have
\begin{eqnarray}
\lim_{\tau^{\prime}\downarrow
0}G_{1}\left(\tau^{\prime}\mid0,\mathbf k
\right)&=&\sum^{\infty}_{l=0}\left(\frac{J(\mathbf
k)}{U}\right)^{l}S_{l} \label{eqnarray 30}
\end{eqnarray}
where
\begin{equation}
S_{l}=\sum^{l}_{k=0}\frac{(l+1)!(l+k)!}{k!(k+1)!(l-k)!\,l!}n^{k+1}
\end{equation}
and $n$ is the particle number on a site. Thus, together with
Eq.(\ref{equation22}), the momentum distribution function
$n(\mathbf k)$ is reduced to
\begin{equation}\label{equation31}
n(\mathbf k)=N_S\mid\emph{w}(\mathbf k)\mid^2\sum^{\infty}_{l=0}
\left(\frac{J}{U}\right)^{l}S_{l}\left[2\cos(k_{x}a)+2\cos(k_{x}\frac
{a}{2}+k_{y}\frac {\sqrt{3}a}{2})+2\cos(k_{y}\frac
{\sqrt{3}a}{2}-k_{x}\frac {a}{2})\right]^{l}.
\end{equation}
This is our analytical result. In the following, we are going to
calculate the time-of-flight absorbtion picture and the visibility
of the interference pattern.

\section{The time-of-flight pictures}

Recently, Becker {\it et al.} \cite{sengstock} performed an
experiment to create an optical triangular lattice and loaded the
ultra-cold $^{87}$Rb atoms in it. The wavelength of the laser
beams which they used to create the lattice is $\lambda=830$ nm,
the corresponding lattice constant is $a=\frac23\lambda$. In fact,
in the experiment, what they created was not a real 2D triangular
lattice but stacked layers of triangular lattices, this stacking
structure is formed by an extra standing wave laser beam. To
eliminate the influence of the third dimension, they set the
lattice depth in the third dimension to be $V_0=30Er$, this setup
makes the tunnelling in the third dimension negligible, and the
system exhibits the property of 2D. In their experiment, they
filled about 40 layers and the occupation for each layer amounts
to about $4000$, together with the chemical potential being
$\mu=133$ nk, a good estimation of the occupation number per site
is $n=2$.

In order to compare our analytical result with their experimental
observation, the third dimension has to be taken into account. Due
to the orthonormality of the Wannier function, the third dimension
would not affect the in-plane hopping parameter, $J$ can be
calculated explicitly from the lattice potential
(\ref{lattice-potential}) and the Wannier function
(\ref{wannierfunction}) via Eq.(\ref{J-formula}), it reads
\begin{equation}
J=Er\left[\frac{\pi^2}{2}-\frac54-\left(\frac1{2\widetilde{V}_0}\right)^{\frac12}\right]\widetilde{V}_0\exp\left\{
-\frac{\pi^2}{3}(2\widetilde{V}_0)^{\frac12}\right\}.
\label{new-J}
\end{equation}
However, the scattering behavior of the system would be 3D, i.e.
in calculating $U$, the interaction strength should be taken as
$g=4\pi \hbar^2 a_s/m$ ($a_s=5.34$ nm is the 3D $s$-wave
scattering length of $^{87}$Rb and $m$ is the corresponding atomic
mass) and the integral of the third dimensional Wannier function
has to be performed accordingly. According to the experiment, by
taking the lattice depth $V_0=30Er$ in the third dimension, a
detailed calculation leads to
\begin{equation}
U=4Er\frac{a_s}{a}(\pi\widetilde{V}_0)^{1/2}(30)^{1/4}.
\label{new-U}
\end{equation}

\begin{figure}[h]
\centering
\includegraphics[width=2.5cm]{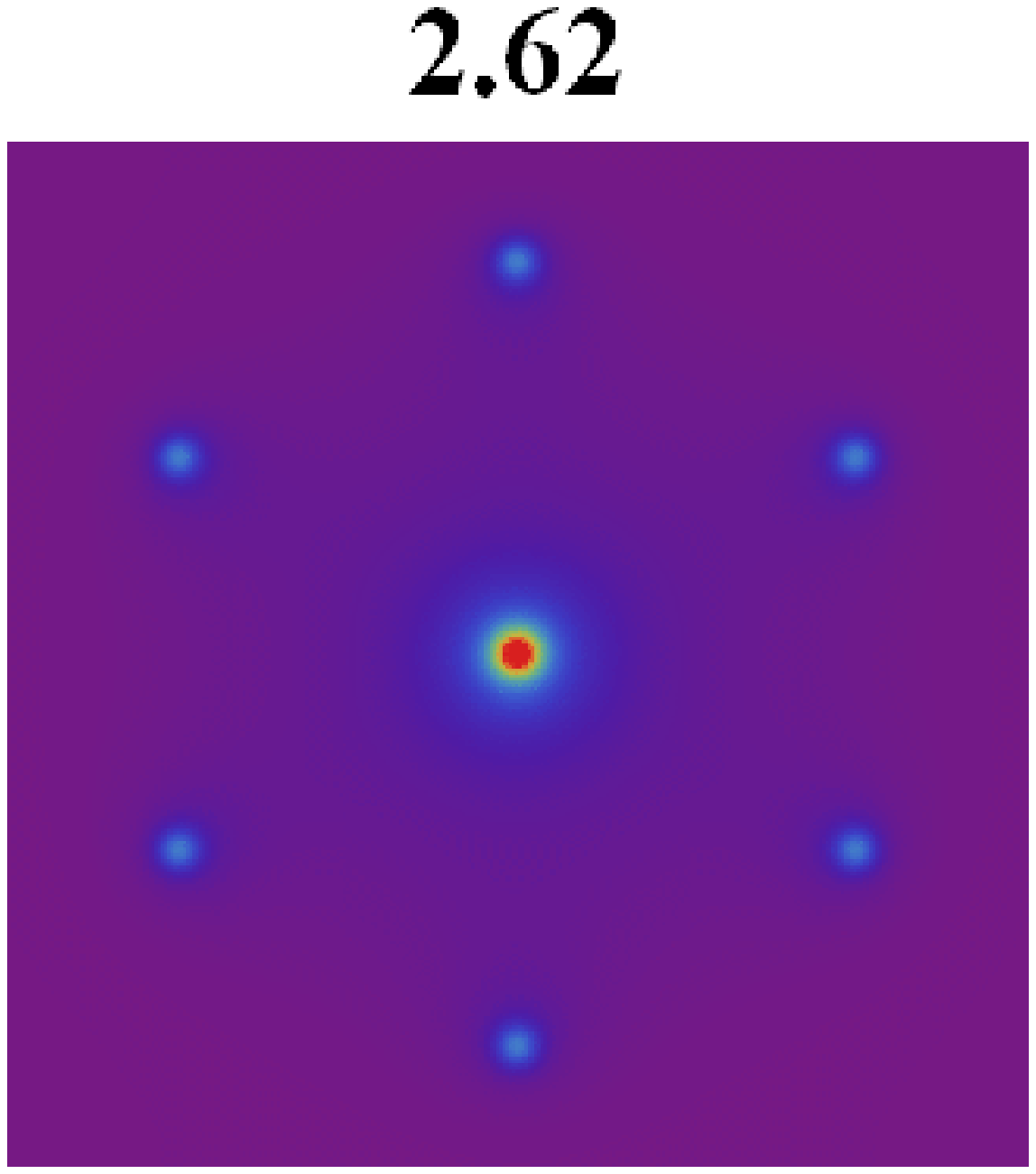}
\includegraphics[width=2.5cm]{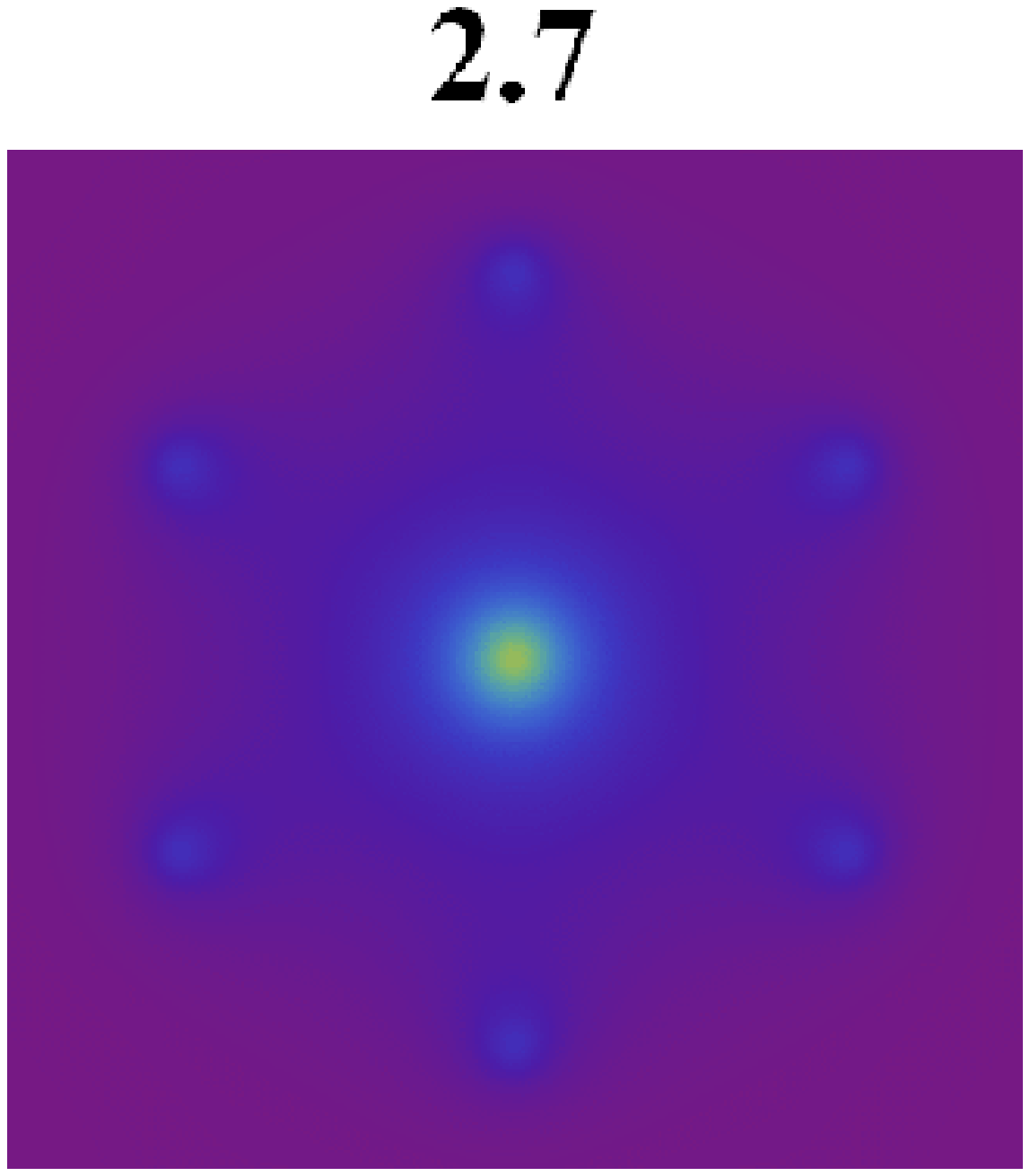}
\includegraphics[width=2.5cm]{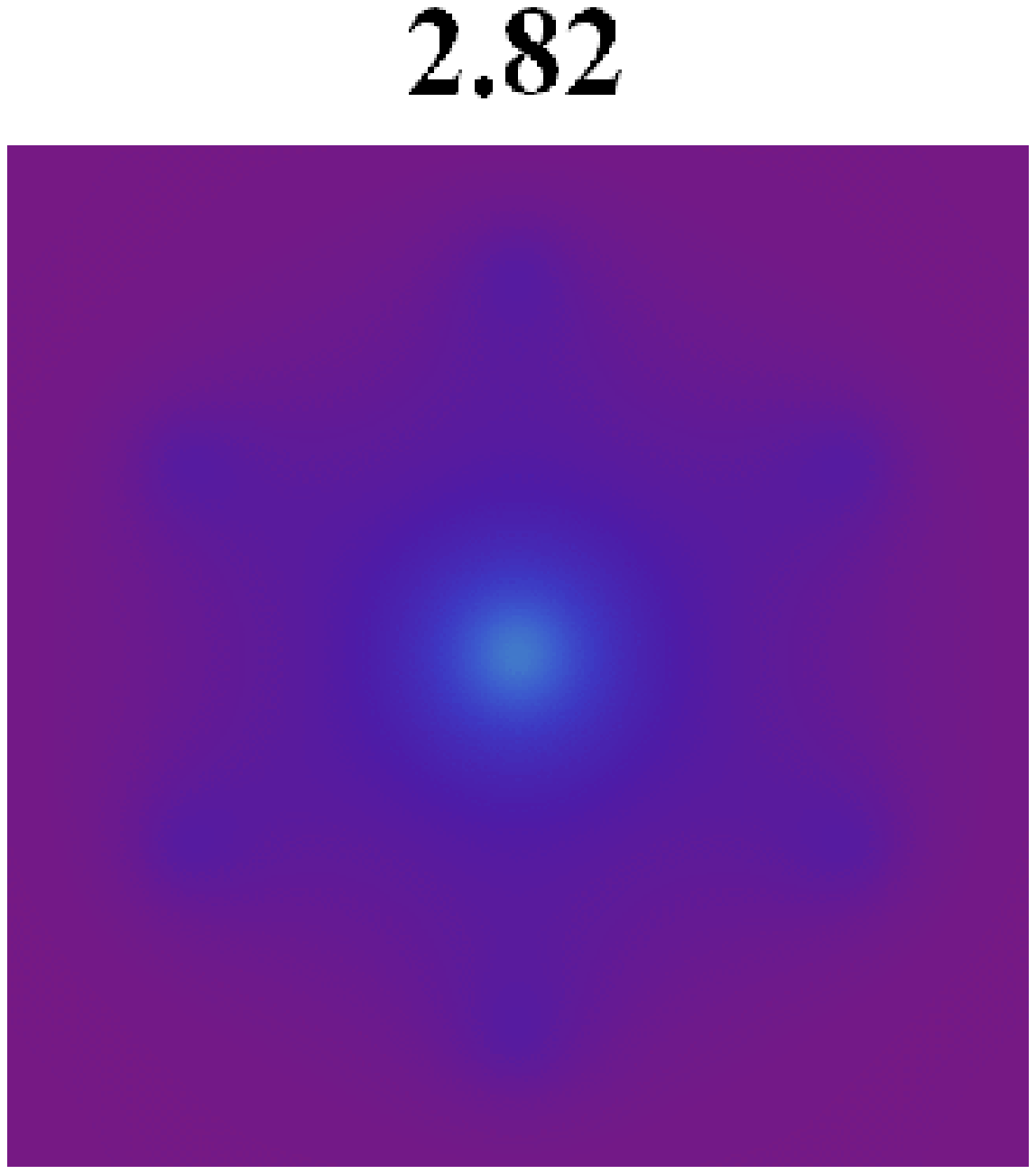}
\includegraphics[width=2.5cm]{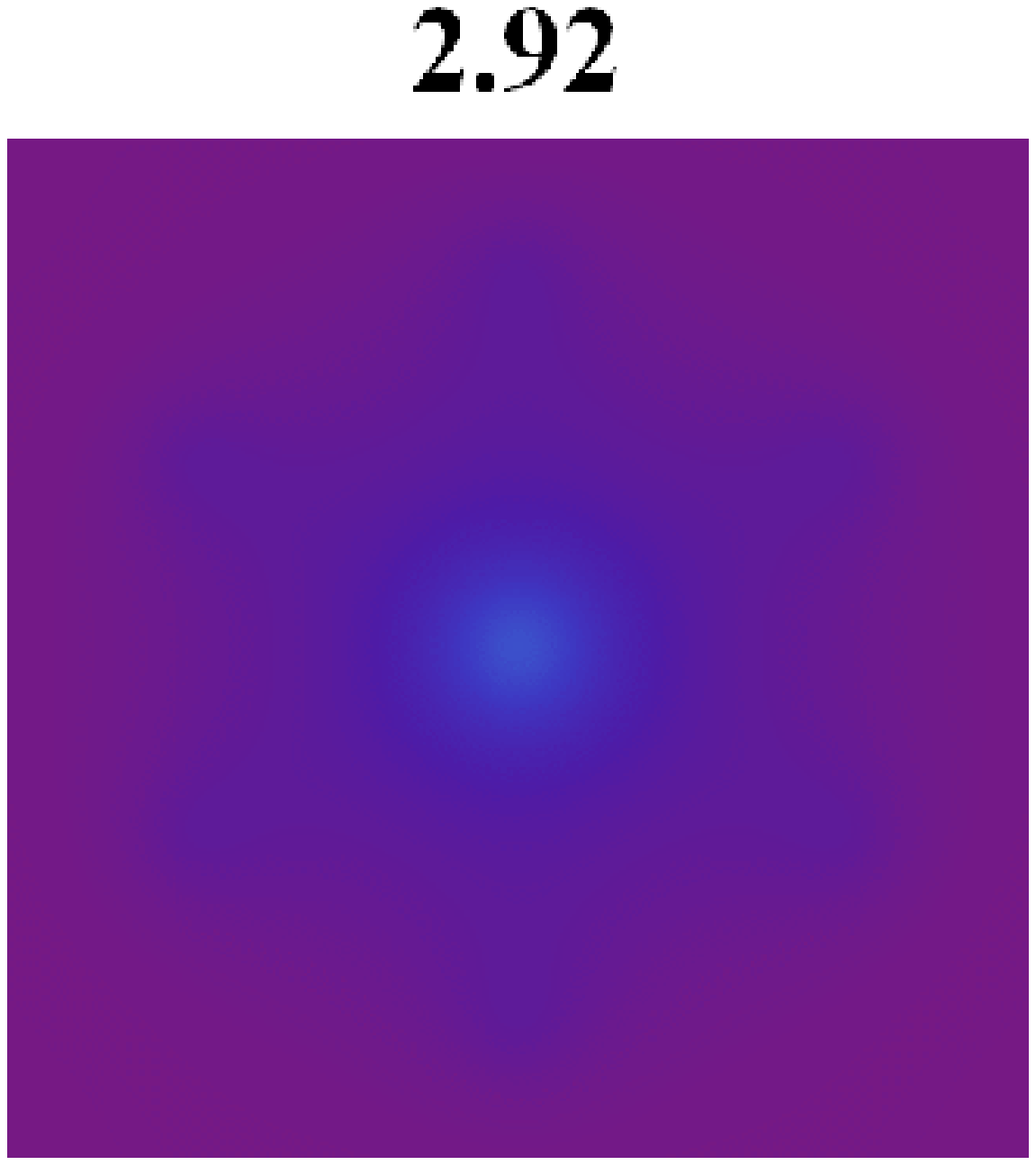}
\includegraphics[width=2.5cm]{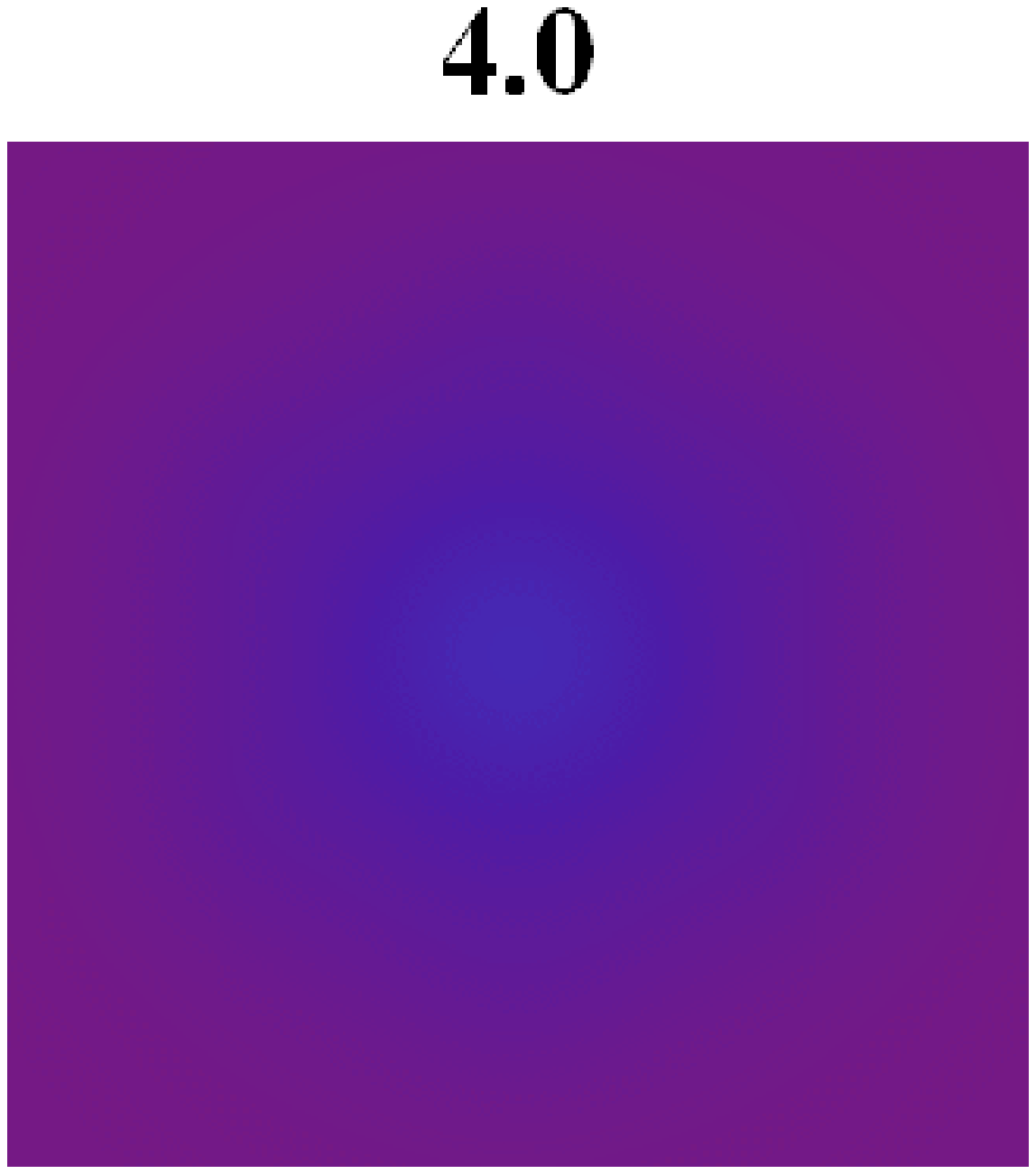}
\caption{The single chain Green's calculation of time-of-flight
absorbtion picture for an ultra-cold Bose system in a triangular
optical lattice for various $\widetilde{V}_0$.}
\label{analytical-picture}
\end{figure}

With (\ref{new-J}) and (\ref{new-U}) as well as the Fourier
transformation of the Wannier function (\ref{wannierfunction}) in
hand, we plot the analytical expression Eq.(\ref{equation31}) of
the momentum distribution function $n({\bf k})$ for various
$\widetilde{V}_0$ in Fig. \ref{analytical-picture}. Qualitatively,
it is in a good agreement with what observed in the experiment
\cite{sengstock}.

\section{The visibility}

\begin{figure}[h!]
\centering \subfigure[]{{\includegraphics[width=7cm]{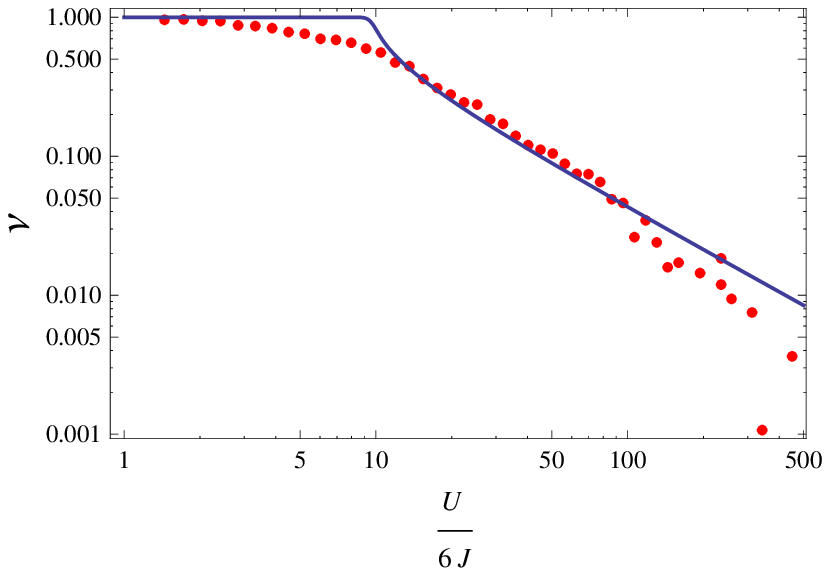}}}
\hspace{0.3in}
\subfigure[]{\raisebox{0.3cm}{{\includegraphics[width=7cm]{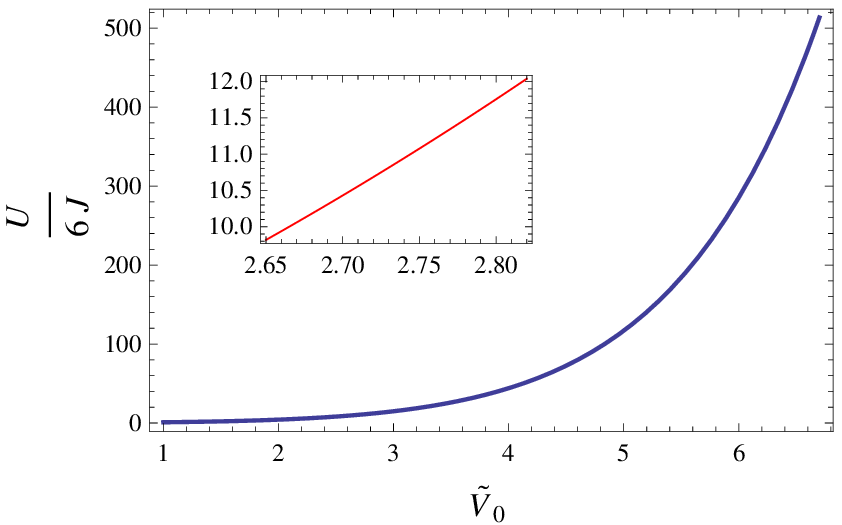}}}}\caption{(a)
The visibility  of the interference pattern against $U/(6J)$. The
solid blue line is the theoretical result from Eq.(\ref{equation
35}) and the red dots are from the experimental observation
\cite{sengstock}. (b) The dependence of $U/(6J)$ on
$\widetilde{V}_0$. } \label{visibility}
\end{figure}

However, the time-of-flight absorbtion picture can only provide us
qualitative impression of the SF-MI phase transition. In addition,
due to various experimental conditions, for instance the
sensitivity of the detectors in the experimental setup, it is hard
to compare the theoretical result of the time-of-flight pictures
directly to experimental ones quantitatively. In order to winkle
quantitative information out of the time-of-flight pictures, the
so-called visibility \cite{Gerbier,Gerbier2} has to be calculated.
The visibility is defined as
\begin{equation}
\nu=\frac{n_{\rm max}-n_{\rm min}}{n_{\rm max}+n_{\rm min}},
\label{equation 35}
\end{equation}
From Eq.(\ref{equation31}), it is easy to find out that $n_{\rm
max}$ takes place at $(0, 4\pi/(\sqrt{3}a))$ and $(2\pi/a,
2\pi/(\sqrt{3}a))$ and other four equivalent points in the
momentum space while $n_{\rm min}$ is at $(2\pi/(\sqrt{3}a),
2\pi/a)$ , $( 4\pi/(\sqrt{3}a),0)$, and other four equivalent
points. All these points have the same distance from the original
point, hence at these points the Wannier function $w({\bf k})$
takes the same value, thus the visibility is solely determined by
the correlations at these points. The theoretical result as well
as the experimental data taken from Ref.[\onlinecite{sengstock}]
are plotted in Fig.\ref{visibility}a.

From Fig.\ref{visibility}a, it is easy to see that our analytical
result shows that the visibility of the interference peaks of the
system drops dramatically fast in the region of
$10\leq\widetilde{V}_{0} \leq12$. In Fig.\ref{visibility}b, We
plot the dependance of $U/(6J)$ on $\widetilde{V}_0$, the
corresponding range of $\widetilde{V}_0$ is from 2.65 to 2.82, as
can be seen in the inset of Fig.\ref{visibility}b. The
experimental estimation is that the corresponding region is from
3.5 to 4. There is roughly a 20\% deviation between our analytical
result and the experimental data, this is understandable: first,
our calculation is a sort of mean-field result; second, during the
calculation of $J$ and $U$, the Gaussian approximation of the
Wannier function is adopted. Nevertheless, our theoretical
prediction is in a reasonable agreement with the experimental
data. Moreover, from the figure of visibility, we see that the
visibility is non-zero far deep inside the Mott state, this is due
to the short-range coherence\cite{Gerbier2} in Mott insulator.

\section{Discussion}

In conclusion, by treating the hopping parameter in Bose-Hubbard
model as a perturbation, with the help of the re-summed Green's
function method and cumulants expansion, the momentum distribution
function of the ultra-cold Bose system in triangular optical
lattice is calculated analytically. By utilizing it, the
time-of-flight absorption picture is plotted and the corresponding
visibility is determined. The comparison between our analytical
results and the experimental data from
Ref.[\onlinecite{sengstock}] exhibits a qualitative agreement.

As we have seen in the discussion of the paper, this systematic
approach can in principle be extended to regime beyond mean-field
theory by adding loop diagrams, more precise expressions of
Wannier function may be adopted to improve the accuracy of the
analytical result. Moreover, the method presented here can not
only be used to investigate homogeneous cold atomic system, but
also can be used to systems with nearest neighbor repulsive
interactions. Ultra-cold Bose gas on a triangular optical lattice
accompanied by nearest neighbor repulsive interaction would
exhibit geometrical frustration effect, hence our present work may
shed some new light in this field.

\section*{Acknowledgement}

We thank C. Becker for providing the experimental data of the
triangle lattice and fruitful discussion. We have also profitted
from stimulating discussion with A. Pelster and F. E. A. dos
Santos. Work supported by Science \& Technology Committee of
Shanghai Municipality under Grant No. 09PJ1404700, and by NSFC
under Grant No. 10845002.


\begin{thebibliography}{99}

\bibitem{Markus-Greiner}M. Greiner, O. Mandel, T. Esslinger, T. W. H\"{a}nsch,
and I. Bloch, Natrue(London) \textbf{415}, 39 (2002).

\bibitem{Immanuel-Bloch}I. Bloch, J. Dalibard, and W. Zwerger,
Rev. Mod. Phys. \textbf{80}, 885 (2008).

\bibitem{lewenstein-1}M. Lewenstein, A. Sanpera, V. Ahufinger, B. Damski,
A. S. De, and U. Sen, Adv. Phys. \textbf{56}, 243 (2007).

\bibitem{sengstock}C. Becker, P. Soltan-Panahi, J. Kronj\"{a}er, S. D\"{o}scher, K. Bongs, and K. Sengstock,
NJP \textbf{12}, 065025 (2010).

\bibitem{sengstock-science}J. Struck, C. \"Olschl\"ager, R. Le
Targat, P. Soltan-Panahi, A. Eckardt, M. Lewenstein, P.
Windpassinger, and K. Sengstock, Science {\bf 333}, 996 (2011)

\bibitem{Eckardt}A. Eckardt, P. Hauke, P. S. Parvis, C. Becker, K. Sengstock and M. Lewenstein,
Europhys. Lett. \textbf{89}, 10010 (2010).

\bibitem{jiang-pra} Z. Lin, J. Zhang, and Y. Jiang, Phys. Rev. A
{\bf 85}, 023619 (2012)

\bibitem{axel}F.E.A. dos Santos and A. Pelster, Phys. Rev. A {\bf 79}, 013614 (2009); B. Bradlyn, F. E. A. dos Santos and A. Pelster,
Phys. Rev. A \textbf{79}, 013615 (2009).

\bibitem{numerical-results} N. Elstner and H. Monien, Phys. Rev. B
{\bf 59}, 12184 (1999); N. Teichmann, D. Hinrichs, and M.
Holthaus, Europhys. Lett. {\bf 91}, 10004 (2010)

\bibitem{Markus-Greiner1}M. Greiner, I. Bloch, O. Mandel, T. W. H\"{a}nsch, and T. Esslinger,
Phys. Rev. Lett. \textbf{87}, 160405 (2001).

\bibitem{Michael}M. K\"{o}hl, H. Moritz, T. St\"{o}ferle, K. G\"unter, and T. Esslinger,
Phys. Rev. Lett. \textbf{94}, 080403 (2005).

\bibitem{Metzner} W. Metzner,
Phys. Rev.B \textbf{43}, 8549 (1993).

\bibitem{ohliger}M. Ohliger, Diploma thesis, Free University of Berlin, 2008, http://users.physik.fu-berlin.de/~ohliger/Diplom.pdf

\bibitem{jaksch-1}D. Jaksch, C. Bruder, J. I. Cirac, C. W. Gardiner, and P. Zoller,
Phys. Rev. Lett. \textbf{81}, 3108 (1998).

\bibitem{blakie}P.B. Blakie, C.W. Clark, J. Phys. B {\bf 37}, 1391
(2004)

\bibitem{prokofev-1}V. A. Kashurnikov, N. V. Prokof¡¯ev, and B. V. Svistunov,
Phys. Rev. A \textbf{66}, 031601(R)(2002).

\bibitem{stoof}D. van Oosten, P. van der Straten, and H.T.C.
Stoof, Phys. Rev. A {\bf 63}, 053601 (2001)

\bibitem{M. Peskin} M. Peskin and D. Schr\"{o}der,{\it An Introduction to Quantum Field Theory} (Westview Press, Boulder, 1995).

\bibitem{Axel2}A. Hoffmann and A. Pelster, Phys.Rev.A \textbf{79}, 053623 (2009).

\bibitem{Gerbier}F. Gerbier, A. Widera, S. F\"{o}ling, O. Mandel, T. Gericke, and I. Bloch,
Phys. Rev. A \textbf{72}, 053606 (2005).

\bibitem{Gerbier2}F. Gerbier, A.Widera, S. F\"{o}ling, O. Mandel, T. Gericke, and I. Bloch,
Phys. Rev. Lett. \textbf{95}, 050404 (2005).




\end{thebibliography}
\end{document}